\documentclass[12pt]{article}
\newcommand{\be}{\begin{equation}}
\newcommand{\ee}{\end{equation}}
\newcommand{\beqn}{\begin{eqnarray}}
\newcommand{\eeqn}{\end{eqnarray}}
\newcommand{\lam}{\lambda}
\newcommand{\eps}{\epsilon}
\def\p{\partial}
\newcommand{\PRL}{Phys. Rev. Lett. }

\newcommand{\PRD}{Phys. Rev. {\bf D}}
\newcommand{\NPB}{Nucl. Phys. {\bf B}}

\newcommand{\f}{\frac}
\newcommand{\w}{\wedge}
\newcommand{\nn}{\nonumber}

\begin{document}

\begin{titlepage}
\begin{center}
\hfill   WIS/13/05-MAY-DPP
\hfill   hep-th/0505257

\vskip 1cm

{\LARGE {\bf $\beta$-deformations, potentials and KK modes         }}

\vskip 1.5cm

{\large  Shesansu Sekhar Pal }\\
\vskip 1.5cm
{ Weizmann institute of  science,\\
76100 Rehovot, Israel \\
\sf shesansu.pal@weizmann.ac.il}

\vskip 3.5cm

\noindent

{\bf Abstract}
\end{center}

We have studied volumes of the 3-cycle  and the compact  5-volumes for 
the $\beta$ transformed geometry and it comes out to be decreasing
except  one  choice  for which
the torus do not stay inside  the 3-cycle and ``5-cycle.'' 
There are 3 possible ways
to construct  these cycles. one is  as mentioned above and the other two are,
when the torus stay inside the cycle and when both the torus and the cycle 
shares a common direction. 
Also, we have argued that under  $\beta$ deformation there arises
a non-trivial ``potential''  as
the $SL(3,R)$ transformation mixes up the fields. If we start with 
a flat space after  the $SL(3,R)$ transformation the Ricci-scalar 
of the transformed geometry do not vanishes but the transformed solution is 
reminiscent of NS5-brane.  We have explicitly, checked that
$\beta$-transformation indeed is a marginal deformation in the gravity side.

\end{titlepage}

\newpage

\section{Introduction}

In the absence of having a technique to solve the sigma model in curved space
and the hope of trying to 
understand the subject as clearly as possible tells us to look for 
various solution to low energy effective theory i.e. the solution 
to supergravity equation of motion. Even though this approach is not
the cleanest one but still can provide us some interesting insights.
The most intriguing aspect is to look for the solution in this kind 
of effective theories which features properties like confinement, chiral
symmetry breaking etc. There is  also a desire to find a dual  pure super 
Yang-Mills theory  which possess these properties at IR. 

Recently there is a technique advocated in \cite{oljm} to construct new 
 solutions from the existing ones. In this study the authors 
of \cite{oljm} 
showed explicitly how to generate new solutions and the interpretation of 
this new solution in the  dual field theory. To summarize their approach, 
they used the global symmetries to construct new solutions. More precisely, 
they
first constructed a torus with two U(1)'s, so these U(1)'s has to be the 
symmetry of the old solution,  and combine these U(1)'s with the SL(2,R) 
symmetry of IIB supergravity. The net result is an SL(3,R) symmetry. Now, 
one can apply this SL(3,R) symmetry on the old solutions to generate new
solution. As mentioned in \cite{oljm} that 
%the solution preserves the same amount of supersymmetry and 
the above procedure can break all supersymmetry 
if at least one of the U(1) coincides with the $U(1)_R$, i.e. if one constructs
a torus by taking one of the direction along the R-symmetry direction then 
the new solution is a non-supersymmetric solution. This procedure is the 
analogue of Leigh-Strassler \cite{ls} deformations in the gravity side 
\cite{oljm}. The marginal deformation to N=4 field theories has been studied 
in \cite{bjl}.  

In the field theory side this SL(3,R) transformed solution corresponds to 
multiplying fields in a different way \cite{oljm}. More explicitly, 
if $\phi_i$ and 
$\phi_j$ are  two chiral super fields with U(1) charges as $q_i$ and $q_j$
then  
\be
\label{product_rule}
\phi_i ~~\phi_j \rightarrow \phi_i \star \phi_j= e^{i\pi \beta {\rm det q}} 
\phi_i ~~
\phi_j,
\ee
where $q$ is a $2\times 2$ matrix and its elements are the $U(1)\times U(1)$ 
charges 
of fields $\phi_i$ and $\phi_j$. This way of deforming the product of 
fields is almost the same as is done in the non-commutative field theories 
\cite{sw}. This is almost because here the $B$ field is not necessarily to 
be a constant.

The two U(1)'s are associated  to the two sides of a torus and  the role of
 modular parameter of the torus is played by the component of the $B$ field
along the torus  and the volume of the torus. 
Consider a geometry which asymptotes to AdS 
spacetime times a compact Sasaki-Einstein manifold. If the torus constructed
above stay inside the AdS space then in the field theory it corresponds to 
a non-commutative field theory with momentum playing the role of the charges
under the U(1)s. Whereas when the torus stay completely inside the SE manifold
then the corresponding fields in the field theory  are multiplied by the 
eq.(\ref{product_rule}). If the torus stay both in the AdS and in the SE space
by sharing one of its direction then  the corresponding field theory in 
\cite{gn} is called as a dipole deformation of the field theory \cite{oj}. 
There are some recent advancement on the study of Lunin-Maldacena background 
\cite{sf}. 

This procedure could be 
useful to generate both the supersymmetric and non-supersymmetric solutions
for the construction of phenomenological models. More interestingly, the 
$\beta$-deformations in the gravitational side can change the  geometry.
What we meant is it can change the values to the coordinate invariant 
quantities and it should as under SL(3,R), which is the $\beta$-deformation
in the gravity, fields gets mix up in a nontrivial way. 
Let us look at 
the simplest example, the flat space, how this technique generates 
non-trivial terms in the action. The flat space example is also studied in 
\cite{oljm}. But before that let us write down the IIB supergravity action 
(in Einstein frame) as 
\be
\label{alpha^'_exp}
S\sim\f{1}{\alpha^{\prime^2}} \bigg (S_1+\alpha^{\prime} S_2+\ldots \bigg )=
\sum^{\infty}_{n=0} {\alpha^{\prime}}^{n-2} S_n,
\ee    
where $S_1=R\pm (\p \Psi)^2$ and $S_2={\cal R}^2 \pm (\p \Psi)^4\pm  (\p^2 \Psi)^2+\ldots$, schematically, with $\Psi$ denotes fields such as dilaton, axion, NSNS and RR 3-form fields and RR 5-form field strength and ${\cal R}^2$ 
denotes ${\cal R}^2=R^2-4 R_{MN}R^{MN}+R_{MNKL} R^{MNKL}$ with $R$ as the 
Ricci scalar. 

Let us start writing down the trivial flat spacetime solution to IIB 
supergravity in the following way
\be
ds^2=e^{-\phi/2}[\eta_{\mu\nu} dx^{\mu\nu}+\sum^3_{i=1} dr^2_i+ r^2_1(d\psi-d\varphi_2)^2+r^2_2(d\psi+d\varphi_1+d\varphi_2)^2+ r^2_3(d\psi-d\varphi_1)^2],
\ee
with all other fields set to zero except dilaton which has been set to a 
constant. For this solution  each $S_n$ in $S$  
vanishes trivially\footnote{We shall be talking to leading order in $\alpha^{\prime}$.}. 

Now, apply the SL(3,R) symmetry, details can be seen in the next section, with 
the following two 1-forms
\be
A^1=-d\psi+\f{3r^2_1r^2_2}{r^2_1r^2_2+r^2_2r^2_3+r^2_3r^2_1}d\psi;\quad A^2=-d\psi+\f{3r^2_3r^2_2}{r^2_1r^2_2+r^2_2r^2_3+r^2_3r^2_1}d\psi.
\ee        
The resulting solution is 
\beqn
\label{ns5}
{ds^{\prime}}^2_E&=&G^{-1/4}[\eta_{\mu\nu} dx^{\mu\nu}+\sum^3_{i=1} dr^2_i]+\nn \\& &G^{3/4}[r^2_1(d\psi-d\varphi_2)^2+r^2_2(d\psi+d\varphi_1+d\varphi_2)^2+ r^2_3(d\psi-d\varphi_1)^2+9\gamma^2 r^2_1r^2_2 r^2_3 d\psi^2]\nn \\
B&=& \gamma G[ r^2_1r^2_2+r^2_2r^2_3+r^2_3r^2_1] D\varphi^1\w D\varphi^2;\quad 
e^{2\phi}=G;\quad G^{-1}=1+ \gamma^2[r^2_1r^2_2+r^2_2r^2_3+r^2_3r^2_1].\nn \\
\eeqn

This solution is reminiscent of the low energy description of  NS5-brane. 
In order to see that we may have 
to S-dualise eq.(\ref{ns5}) and consider some choice to $\gamma$.  
It would  be interesting to quantize strings in 
this background.  

The first term of $S_1$ in eq.(\ref{alpha^'_exp}) is
\beqn
S_1&=&\sqrt g^{\prime}_E R^{\prime}_E; \quad \sqrt g^{\prime}_E= 3r_1 r_2 r_3 G^{-1/4}\nn \\
R^{\prime}_E&=&\f{\gamma^2[4(r^2_1+r^2_2+r^2_3)+\gamma^2(3r^4_1(r^2_2+r^2_3)+3r^4_2(r^2_3+r^2_1)+3r^4_3(r^2_1+r^2_2))+6\gamma^2r^2_1r^2_2r^2_3]}{2[1+\gamma^2(r^2_1r^2_2+r^2_2r^2_3+r^2_3r^2_1)]^{9/4}},\nn \\ 
\eeqn
non-zero ( in the solution we have kept one parameter $\gamma$). 
%But, the total action in that order of $\alpha^{\prime}$ has to 
%vanish because we started out with a flat spacetime and SL(3,R) can't change
%the value of the action i.e. $S_1$. which means that 
However, if we write $S_n=T_n-V_n$, after compactification then 
the  potential $V_n$ has been changed from zero to non-zero due to SL(3,R)
transformation. 

Applying this procedure to some non-trivial examples like KS solution 
\cite{ks}, it seems
that we can generate some non-trivial potential.

%It also seems that applying this technique we may not  be able to see the 
%divergence
%in the coordinate invariant quantities like $R, R_{MN}R^{MN}, 
%R_{MNKL}R^{MNKL}$  in the  $\beta$-deformed solution. 
%Hence, this method could be useful for the removal
%of the singularity as it mixes the fields in  a non-trivial way.  

%to check the regularity of the SL(3,R)
%transformed solution it is important to check the regularity of $S_n$ for 
%each $n$, not just the coordinate invariant quantities like $R, R_{MN}R^{MN}, 
%R_{MNKL}R^{MNKL}$.       

In this paper, we have just argued how to generate non-trivial potential in
the effective theory using the prescription of \cite{oljm}. The example we 
choose  was  a simple one, however it would be interesting to apply this 
technique for some interesting geometries and see the consequences of it. 
We shall write down the most general solution in IIB supergravity
following \cite{oljm} in section
2 and in section 3, we shall calculate the volumes of various 3-cycle. 
There are three ways to form a 3-cycle  depending on the location of the 
torus: torus staying inside the 3-cycle, the torus staying outside the 
3-cycle and
if the torus shares one of its direction with the 3-cycle. 
However, computation of these objects for a general 
$\beta$-deformed background is not that simple, so what we do is to compute
it for a specific choice of the background and show that when does the volumes 
decreases and increases. Since, the masses of the KK modes are related to 
these volumes means we got a feeling of when can we  possibly be able to
integrate out  these KK modes, for what choice of cycles, so as to obtain a 
pure super Yang-Mills in the confining theory. In section 4, we 
compute the sizes of the compact 5-manifold depending on the choice position 
of the torus, following the same strategy as in section 3. 
%which is important for the construction of the $\beta$-deformation. 
%This technique we employed also in section 3. 
In section 5, we show explicitly
why this is called a marginal deformation by computing the central charge
and the R-charges which is related to the dimension of the chiral super fields.
We compute the tension of the fundamental string in the $\beta$-deformed
solution, which shows that in general the tensions are not same and 
conclude in the end.

\section{The solution}
The most general metric for IIB supergravity that can be written with
explicit  $U(1)\times U(1)$  symmetry is \cite{oljm}\footnote{Of course other 
symmetries like SL(2,R), Poincare symmetry etc. are there.}:

\beqn 
\label{gen_sol} 
ds^2_{IIB}&=&F  \bigg[ \f {1}
{\sqrt\Delta}  (D\varphi^1-C D\varphi^2)^2 +{\sqrt\Delta}
(D\varphi^2)^2 \bigg]
+\f{e^{2\phi/3}}{F^{1/3}}g_{\mu\nu}dx^{\mu} dx^{\nu},\nn \\
B&=&B_{12}D\varphi^1 \wedge D\varphi^2+\bigg[
B_{1\mu}D\varphi^1+B_{2\mu}D\varphi^2\bigg]\wedge dx^{\mu}-
\f{1}{2}( A^m_{\mu}B_{m\nu}-{\tilde b}_{\mu\nu})dx^{\mu}\wedge dx^{\nu} \nn \\
& &e^{2\Phi}=e^{2\phi},\quad C^{(0)}=\chi \nn \\
C_{(2)}&=& C_{12}D\varphi^1 \wedge D\varphi^2+\bigg[
C_{1\mu}D\varphi^1+C_{2\mu}D\varphi^2\bigg]\wedge dx^{\mu}-
\f{1}{2}( A^m_{\mu}C_{m\nu}-{\tilde c}_{\mu\nu})dx^{\mu}\wedge dx^{\nu}
\nonumber\\
C_{(4)}&=&-\frac{1}{2}(
{\tilde d}_{\mu\nu}+B_{12}{\tilde c}_{\mu\nu}-
\eps^{mn}B_{m \mu}C_{n\nu}-B_{12}A^m_{\mu}C_{m\nu})
dx^\mu \w dx^\nu\w D\varphi^1\w D\varphi^2 \nonumber\\
& &+\frac{1}{6}[{  C}_{\mu\nu\lam}+3({\tilde b}_{\mu\nu}+
A^1_{\mu}B_{1\nu}-A^2_{\mu}B_{2\nu})C_{1\lam}] dx^{\mu}\w dx^{\nu}\w
dx^\lam \w D\varphi^1 + 
\\ && + d_{\mu_1 \mu_2\mu_3\mu_4} dx^{\mu_1} \w dx^{\mu_2}\w dx^{\mu_3}\w dx^{\mu_4} +
 \hat d_{\mu_1 \mu_2 \mu_3  } dx^{\mu_1} \w dx^{\mu_2} \w dx^{\mu_3}\w D\varphi^2
\nonumber 
\eeqn 
where 
\beqn
D\varphi^1\equiv d\varphi^1+A^1_{\mu} dx^{\mu},\qquad D\varphi^2\equiv d\varphi^2+{A}^2_{\mu} dx^{\mu} 
\eeqn
and the  coefficients like $ d_{\mu_1 \mu_2\mu_3\mu_4}$ and 
$ {\hat d_{\mu_1 \mu_2 \mu_3}} $
 has to be determined from the self duality 
of ${\tilde F}_5$ i.e. ${\tilde F}_5=F_5+\star_{10}F_5$  with $F_5=dC_4$. 
Also, the
coefficients that appeared in eq.(\ref{gen_sol}) are independent of 
$\varphi^1, 
\varphi^2$ coordinates i.e. they are functions of $x_{\mu}$ only. The U(1)'s
act by shifting the $\varphi^1, \varphi^2$ coordinates. 
Here $\mu,\nu=1,\cdots,8$  and $m,n=1,2$.

The action of $SL(3,R)$ transformations are : 
\beqn
&&V^{(1)}_\mu=\left(\begin{array}{c} -B_{2\mu}\\A^1_\mu\\C_{2\mu}
\end{array}\right),\quad
V^{(2)}_\mu=\left(\begin{array}{c}
B_{1\mu}\\A^2_\mu\\-C_{1\mu}
\end{array}\right):\quad V^{(i)}_\mu\rightarrow (\Lambda^T)^{-1}V^{(i)}_\mu;
\label{1forms}\\
&&
W_{\mu\nu}=\left(\begin{array}{c}
{\tilde c}_{\mu\nu}\\{\tilde d}_{\mu\nu}\\
{\tilde b}_{\mu\nu}
\end{array}\right)\rightarrow \Lambda W_{\mu\nu}
\label{2forms}
\eeqn
and defining a matrix $M=g g^T$ with 
\beqn
g^T=
\left(\begin{array}{ccc}
e^{-\phi/3}F^{-1/3}&0&0\\
0&e^{-\phi/3}F^{2/3}&0\\
0&0&e^{2\phi/3}F^{-1/3}
\end{array}\right)
\left(\begin{array}{ccc}
1&B_{12}&0\\
0&1&0\\
\chi&-C_{12}+\chi B_{12}&1
\end{array}\right) 
\label{demag}
\eeqn 
which transforms as \beqn 
M\rightarrow \Lambda  M\Lambda^T
\eeqn
 
The scalars $\Delta, ~C$ as well as the three form ${
C}_{\mu\nu\lam}$ do not changes  under these $SL(3,R)$
transformations. The transformed form of  $  
d_{\mu_1 \mu_2\mu_3\mu_4}, {\hat d_{\mu_1 \mu_2\mu_3}}$ that appear in
$C^{(4)}$ has to be determined by imposing the self duality condition 
 on the five form field strength.
The metric $g_{\mu \nu}$ is the Einstein metric in eight
dimensions and does not change under any of these transformations.

The 2-parameter choice of $\Lambda$ which is an element of SL(3,R)  is: 

\beqn 
\label{lam_mat}
\Lambda=\left(\begin{array}{ccc}
1& \gamma &0\\
0& 1& 0 \\
0&\sigma &1
\end{array}\right), \quad 
\Lambda^T=\left(\begin{array}{ccc}
1& 0&0\\
 \gamma&1& \sigma \\
0&0 &1
\end{array}\right), \quad (\Lambda^T)^{-1}=
\left(\begin{array}{ccc}
1& 0&0\\
 -\gamma&1& -\sigma \\
0&0 &1
\end{array}\right).
\eeqn

The transformed matrix $\ M ~~~~{\stackrel {SL(3,R)}{\rightarrow}}
~~~~~~\Lambda g g^T \Lambda^T \ $  is
\beqn
\label{M_matrix}
M^{\prime}=\left(\begin{array}{ccc}
\ell_1& \ell_2 & \ell_3\\
\ell_2& \ell_4& \ell_5 \\
\ell_3&\ell_5 &\ell_6
\end{array}\right),
\eeqn

where
\beqn
\ell_1&=&e^{-2\phi/3} F^{-2/3} (1+\gamma B_{12})^2+\gamma^2 e^{-2\phi/3} 
F^{4/3}+\nn \\& &e^{4\phi/3} F^{-2/3}[\chi+\gamma(-C_{12}+\chi B_{12})]^2\nn \\
\ell_2&=& e^{-2\phi/3} F^{-2/3}B_{12}(1+\gamma B_{12})+\gamma e^{-2\phi/3} 
F^{4/3}+\nn \\& &e^{4\phi/3} F^{-2/3}(-C_{12}+\chi B_{12})[\chi+\gamma(-C_{12}+\chi B_{12})]\nn \\
\ell_3&=&\sigma e^{-2\phi/3} F^{-2/3}B_{12}(1+\gamma B_{12})+\sigma \gamma e^{-2\phi/3} 
F^{4/3}+\nn \\& &e^{4\phi/3} F^{-2/3}[1+\sigma (-C_{12}+\chi B_{12})]~[\chi+\gamma(-C_{12}+\chi B_{12})]\nn \\
\ell_4&=&B^2_{12} e^{-2\phi/3} F^{-2/3}+ e^{-2\phi/3} F^{4/3}+e^{4\phi/3} F^{-2/3}[-C_{12}+\chi B_{12}]^2\nn \\
\ell_5&=&\sigma B^2_{12} e^{-2\phi/3} F^{-2/3}+ \sigma e^{-2\phi/3} F^{4/3}+\nn \\& &e^{4\phi/3} F^{-2/3}(-C_{12}+\chi B_{12})[1+\sigma(-C_{12}+\chi B_{12})]\nn \\
\ell_6&=&\sigma^2 B^2_{12} e^{-2\phi/3} F^{-2/3}+ \sigma^2 e^{-2\phi/3} F^{4/3}+\nn \\& &e^{4\phi/3} F^{-2/3}[1+\sigma(-C_{12}+\chi B_{12})]^2.
\eeqn

Note that the matrix $M$ is a symmetric matrix, so it has got six independent 
components. However, eq.(\ref{demag}) contains five independent functions, 
which means not all the $\ell_i$'s  are independent. The relation is 
\be
\ell_4=\f{1}{\ell_1\ell_6-\ell^2_3}+\f{\ell^2_5}{\ell_6}+
\f{(\ell_2\ell_6-\ell_3\ell_5)^2}{\ell_6(\ell_1\ell_6-\ell^2_3)}.
\ee

From this matrix $M^{\prime}$ we can read out the various transformed fields.
Those are 
\beqn
e^{2\phi^{\prime}}&=&\f{\ell^2_6}{\ell_1\ell_6-\ell^2_3},\quad 
\chi^{\prime}=\f{\ell_3}{\ell_6},\quad F^{\prime}=\f{\sqrt{\ell_6}}{\ell_1\ell_6-\ell^2_3}\nn \\
B^{\prime}_{12}&=&\f{\ell_2\ell_6-\ell_3\ell_5}{\ell_1\ell_6-\ell^2_3}\nn \\
C^{\prime}_{12}&=&\f{\ell_2\ell_3\ell_6-\ell^2_3\ell_5}{\ell_1\ell^2_6-\ell^2_3\ell_6}-\f{\ell_5}{\ell_6}.\nn \\
\eeqn 

For completeness we write down the transformed ``vector'' and ``tensor'' fields
\beqn
\label{prime_vecs}
& &\left(\begin{array}{c} -B_{2\mu}\\A^1_\mu\\C_{2\mu}
\end{array}\right)\rightarrow \left(\begin{array}{c} -B_{2\mu}\\\gamma 
B_{2\mu}+A^1_\mu-\sigma~ C_{2\mu}\\C_{2\mu}
\end{array}\right), \quad
\left(\begin{array}{c}
B_{1\mu}\\A^2_\mu\\-C_{1\mu}
\end{array}\right)\rightarrow \left(\begin{array}{c}
 B_{1\mu}\\-\gamma B_{1\mu}+A^2_\mu+\sigma C_{1\mu}\\-C_{1\mu}
\end{array}\right)\nn \\
& &\left(\begin{array}{c}
{\tilde c}_{\mu\nu}\\{\tilde d}_{\mu\nu}\\
{\tilde b}_{\mu\nu}
\end{array}\right)\rightarrow \left(\begin{array}{c}
{\tilde c}_{\mu\nu}+\gamma {\tilde d}_{\mu\nu}\\{\tilde d}_{\mu\nu}\\
{\tilde b}_{\mu\nu}+\sigma {\tilde d}_{\mu\nu}
\end{array}\right).
\eeqn

From the transformed value to $A^1$ and $A^2$, it follows that for spacetime
with geometry of the form $AdS_5 \times X_5$ the transformed geometry takes
a rather simple form i.e. if the original geometry had no $B_2$ and $C_2$ 
fields then these two fields won't appear in the  $\beta$-transformed geometry.

 The transformed geometry in string frame is
\be
\label{trans_sol}
ds^2=\f{\sqrt{\ell_6}}{\ell_1\ell_6-\ell^2_3}\bigg[ \f {1}
{\sqrt\Delta}  (D\Phi^1-C D\Phi^2)^2 +{\sqrt\Delta}
(D\Phi^2)^2 \bigg]+{\sqrt{\ell_6}} g_{\mu\nu}
 dx^{\mu} dx^{\nu},
\ee
where
\beqn
D\Phi^1&=&d\varphi^1+(\gamma B_{2\mu}+(A^1)_{\mu}-\sigma C_{2\mu})dx^{\mu}=D\varphi^1+(\gamma B_{2\mu}-\sigma C_{2\mu})dx^{\mu}
\nn \\
D\Phi^2&=&d\varphi^2+(-\gamma B_{1\mu}+(A^2)_{\mu}+\sigma C_{1\mu}) dx^{\mu}=D\varphi^2+(-\gamma B_{1\mu}+\sigma C_{1\mu})dx^{\mu}.
\eeqn

\section{Various cycles}

As we know the desire to construct a pure super Yang-Mills theory in a 
confining theory at the IR is hampered due to the presence of KK modes, whose
masses are of the same order as that of the scale of the field theory. Hence 
its very difficult to decouple them and generate a pure SYMs theory. 

Since, SL(3,R) symmetry mixes up the fields, which imply it could be possible 
to change the masses of KK modes. However, the SL(3,R) symmetry in the dual
field theory has been considered as a marginal deformations in \cite{oljm}
implies the physically interesting objects like R-charge and central charges
of chiral super fields should not change.

There has been a computation of masses of KK modes  in \cite{gn} in 
the $\beta$-deformed MN solution \cite{mn} for a specific choice of cycle.    
The result of \cite{gn} is that the masses of KK modes increases. However, 
we shall try to calculate these objects in general. It seems 
that there is an urgency to construct the supersymmetric cycles in the 
$\beta$-transformed geometry so that  we can construct a pure super
Yang-Mills theory.  

Had it been a relevant deformation then 
we could have said that the masses of the KK modes should decrease under
the assumption that the flow is driven  only by the superpotential coupling.

In any case we shall investigate and consider various cases and try to find
any case, if it exists, such that the masses of KK modes  do not changes. 
However, it does not looks like that would be the  case. We do not know
of any general argument which can tell us about the masses of KK modes but
several examples suggests that 
the masses changes   
at IR. The transformed geometry  depends on a factor, G,  which
is different from identity for non-zero $\gamma$,  one of the parameter which
causes deformation to the geometry and this is the only parameter allowed to 
have a non-singular solution in the interesting cases like KS \cite{oljm}. 

Before we start doing the computation, let us rewrite the geometry in 
eq.(\ref{gen_sol}) and eq.(\ref{trans_sol}) in Einstein frame as
we are going to compute the volumes in this frame 
\beqn
\label{e_frame}
ds^2_E&=&e^{-\phi/2}F  \bigg[ \f {1}
{\sqrt\Delta}  (D\varphi^1-C D\varphi^2)^2 +{\sqrt\Delta}
(D\varphi^2)^2 \bigg]
+\f{e^{\phi/6}}{F^{1/3}}g_{\mu\nu}dx^{\mu} dx^{\nu}\nn \\
ds^{{\prime}^2}_E&=&\f{1}{(\ell_1\ell_6-\ell^2_3)^{3/4}}\bigg[ \f {1}
{\sqrt\Delta}  (D\Phi^1-C D\Phi^2)^2 +{\sqrt\Delta}
(D\Phi^2)^2 \bigg]+(\ell_1\ell_6-\ell^2_3)^{1/4} g_{\mu\nu}\nn \\
 dx^{\mu} dx^{\nu}.
\eeqn

Let us define a quantity 
\be
G\equiv e^{-2\phi}\f{\ell^2_6}{\ell_1\ell_6-\ell^2_3}=
e^{2(\phi^{\prime}-\phi)}, 
\ee
which in the limit of $\chi=0, B_{12}=0, C_{12}=0, \sigma=0$ gives 
$G^{-1}\rightarrow 1+\gamma^2 F^2$. 

Let us denote the torus i.e. the $U(1)\times U(1)$ directions as T and the
directions associated to 3-cycle as ${\cal C}$. So, with this choice there 
exists three ways to construct 3-cycle.  (1) when the two of the directions 
of torus (T) stay inside the 3-cycle    ${\cal C}$, i.e. $T~ || ~{\cal C}$ (2)
when they are orthogonal  $T \perp {\cal C}$  i.e. torus do not stay inside
the $ {\cal C}$ (3) when one of the torus direction stay inside the 3-cycle
i.e. $T \cap {\cal C}=1$. It could be  possible that some choices are not 
realisable in practice. But, we are not worried about that.  \\

{\underline{ Choice 1:   $T~ || ~{\cal C}$}}\\

For this case let us take a choice $x^1\neq 0$ such that the ten dim 
geometry  reduces to a 3-dim geometry i.e.  $x^2=0\cdots 0=x^8$, 
So, the metric components  are functions 
of only  $x^1$ coordinates.  
\beqn
ds^2_E&\rightarrow&e^{-\phi/2}F  \bigg[ \f {1}
{\sqrt\Delta}  (D\varphi^1-C D\varphi^2)^2 +{\sqrt\Delta}
(D\varphi^2)^2 \bigg]
+\f{e^{\phi/6}}{F^{1/3}}g_{11}dx^{1} dx^{1}\nn \\
ds^{{\prime}^2}_E&\rightarrow&\f{1}{(\ell_1\ell_6-\ell^2_3)^{3/4}}\bigg[ \f {1}
{\sqrt\Delta}  (D\Phi^1-C D\Phi^2)^2 +{\sqrt\Delta}
(D\Phi^2)^2 \bigg]+(\ell_1\ell_6-\ell^2_3)^{1/4} g_{11}
 dx^{1} dx^{1}.
\eeqn
 
The volumes are
\beqn
V^{(1)}_3&=&(2\pi)^2 \int e^{-5\phi/12} F^{5/6}{\sqrt g_{11}} dx^1\nn \\
V^{\prime^{(1)}}_3&=& (2\pi)^2 \int \f{{\sqrt g_{11}}}{(\ell_1\ell_6-\ell^2_3)^{5/8}}
dx^1.
\eeqn

In general the integrand of $V^{\prime}_3$ depend on the parameter $\gamma$ 
and $\sigma$, But, 
for the most simplest choice i.e. 
\be
\label{choice}
\chi=0, B_{12}=0, C_{12}=0, \sigma=0, 
\ee
$V^{\prime}_3$ contains the parameter $\gamma$ through $\ell_1\ell_6-\ell^2_3=
e^{2\phi/3} F^{-4/3} (1+\gamma^2 F^2)$. So, for  $T~ || ~{\cal C}$ case the
volumes are not same. 

Now, if we had chosen $x^2$, another coordinate, instead of $x^1$ in  the 
construction of volume 
of the 3-cycle in the deformed geometry, by thinking that  it is this 
coordinates $\varphi^1,\varphi^2,x^2$ that forms a 3-cycle in the deformed 
geometry rather than the one in the undeformed theory, 
then still the deformed volume would
depend on the parameter $\gamma,\sigma$. Hence, for this choice of 3-cycle
there does not seems that volume of it will be unchanged. However, we do say 
that $V^{\prime}_3 ~<~  V_3$, as 
\be
\f{dV^{\prime^{(1)}}_3}{dV^{(1)}_3}=\f{e^{5\phi/12} F^{-5/6}}{(\ell_1\ell_6-\ell^2_3)^{5/8}},
\ee
where $dV^{(1)}_3$ stands for the integrand of $V^{(1)}_3$ which in the limit 
eq.(\ref{choice}) becomes
\be
\f{dV^{\prime^{(1)}}_3}{dV^{(1)}_3}=\f{e^{-\phi/4} F^{1/2}}{1+\gamma^2 F^2}.
\ee
The appearance of $\gamma^2$ in the denominator implies that 
$\f{dV^{\prime^{(1)}}_3}{dV^{(1)}_3} < 1$ and to show that  
$V^{\prime^{(1)}}_3 ~<~  V^{(1)}_3$, we relied  an example of 
$AdS_5\times M^5$.  \\

It could have been as well possible that the 3-cycles in the undeformed theory 
stay in the first choice but the 3-cycles in the deformed theory could stay 
in the second or the third choices, i.e. the 3-cycles both in the deformed and undeformed theory should not necessarily stay in the same choice as assumed 
earlier.\\

{\underline{ Choice 2:   $T~ \perp ~{\cal C}$}}\\
      
In this case proceeding in the same way like the first choice and considering 
that the 3-cycle is defined for 
\be
\varphi^1=0=\varphi^2; \qquad  x^4=0=\cdots=x^8, 
\ee
then we find the metric
from eq.(\ref{e_frame}) as
\beqn
\label{choice_2}
ds^2_E&\rightarrow&\sum^3_{\mu,\nu=1}\bigg[e^{-\phi/2}F\bigg(\f{1}{\sqrt\Delta}(A^1_{\mu}-CA^2_{\mu})(A^1_{\nu}-CA^2_{\nu})+{\sqrt\Delta}A^2_{\mu}A^2_{\nu}\bigg)+\f{e^{\phi/6}}{F^{1/3}}g_{\mu\nu}\bigg]dx^{\mu} 
dx^{\nu}\nn \\
&\equiv & G_{\mu\nu}dx^{\mu} dx^{\nu}\nn \\
ds^{{\prime}^2}_E&\rightarrow&\sum^3_{\mu,\nu=1}\bigg[\f{1}{(\ell_1\ell_6-\ell^2_3)^{3/4}}\bigg(\f{1}{\sqrt\Delta}{\cal A}_{\mu}{\cal A}_{\nu}+{\sqrt\Delta}(-\gamma B_{1\mu}+A^2_{\mu}+\sigma C_{1\mu})(-\gamma B_{1\nu}+A^2_{\nu}+\sigma C_{1\nu}) \bigg)+\nn \\& &(\ell_1\ell_6-\ell^2_3)^{1/4} g_{\mu\nu}\bigg]dx^{\mu} dx^{\nu}\nn \\
&\equiv & G^{\prime}_{\mu\nu}dx^{\mu} dx^{\nu},
\eeqn 
where ${\cal A}_{\mu}=\gamma B_{2\mu}+A^1_{\mu}-\sigma C_{2\mu}+
\gamma CB_{1\mu}-CA^2_{\mu}-
C\sigma C_{1\mu}$. The metric components are functions of $x^1,x^2$ and $x^3$ 
coordinates. The volumes are
\beqn
\label{vol_2}
V^{(2)}_3&=&\int dx^1 dx^2 dx^3 {\sqrt {det G}}\nn \\
V^{\prime^{(2)}}_3&=& \int dx^1 dx^2 dx^3  {\sqrt {det G^{\prime}}}.
\eeqn

As before the $V^{\prime^{(2)}}_3$ depends on the parameters $\gamma,\sigma$
which means $\f{V^{\prime^{(2)}}_3}{V^{(2)}_3}\neq 1$ and in the obvious case 
that is when there is no deformation  $\gamma=0=\sigma$ the ratio 
$\f{V^{\prime^{(2)}}_3}{V^{(2)}_3}= 1$. This is to make sure that we did not
do some error in the computation of the eq.(\ref{vol_2}).\\

%One can see that 
%$\f{dV^{\prime^{(2)}}_3}{dV^{(2)}_3} >1 $ in the limit  eq.(\ref{choice}).\\

{\underline{ Choice 3:   $T~ \cap~{\cal C}=1$}}\

For this case we can have two possibilities depending on the choice to include
either $\varphi^1$ or $\varphi^2$ in the construction of 3-cycle. Since, these 
coordinates do not appear symmetrically in the geometry, so we have to 
evaluate the volumes separately. 

Let us take the choice that the 3-cycle is extended along $\varphi^1, x^1$ and
$x^2$ directions and we shall set $\varphi^2=0=x^3=0=\cdots 0=x^8$. In this 
case the geometries are 
\beqn
\label{varphi2=0}
ds^2_E&\rightarrow&e^{-\phi/2}F\f{1}{\sqrt\Delta}[(d\varphi^1)^2+2d\varphi^1 
 (A^1_{\mu}-CA^2_{\mu})dx^{\mu}]+\nn \\ & &\sum^2_{\mu,\nu=1}\bigg[e^{-\phi/2}F\f{1}{\sqrt\Delta} (A^1_{\mu}-CA^2_{\mu})(A^1_{\nu}-CA^2_{\nu})
+e^{-\phi/2}F {\sqrt\Delta} A^2_{\mu} A^2_{\nu}+
\f{e^{\phi/6}}{F^{1/3}}g_{\mu\nu}\bigg]dx^{\mu} 
dx^{\nu}\nn \\
ds^{{\prime}^2}_E&\rightarrow&\f{1}{(\ell_1\ell_6-\ell^2_3)^{3/4}{\sqrt\Delta}}[(d\varphi^1)^2+2d\varphi^1 
 {\cal A}_{\mu}dx^{\mu}]+\nn \\& &\sum^2_{\mu,\nu=1}\bigg[\f{1}{(\ell_1\ell_6-\ell^2_3)^{3/4}}\bigg(\f{1}{\sqrt\Delta}{\cal A}_{\mu}{\cal A}_{\nu}+{\sqrt\Delta}(-\gamma B_{1\mu}+A^2_{\mu}+\sigma C_{1\mu})(-\gamma B_{1\nu}+A^2_{\nu}+\sigma C_{1\nu}) \bigg)+\nn \\& &(\ell_1\ell_6-\ell^2_3)^{1/4} g_{\mu\nu}\bigg]dx^{\mu} dx^{\nu}\nn \\
\eeqn 

It is not illuminating to compute the volume from the geometries as we can't
compare them. Nevertheless, as we can see that the ratio is 
different from identity which means the KK masses  do changes after the
$\beta$-deformation. 

 Let us write down the geometries 
for the other case i.e. 
$\varphi^1=0=x^3=0=\cdots 0=x^8$, for completeness 

\beqn
\label{varphi1=0}
ds^2_E&\rightarrow&e^{-\phi/2}F[(\f{C^2}{\sqrt\Delta}+{\sqrt\Delta})(d\varphi^2)^2+2d\varphi^2 dx^{\mu}
((\f{C^2}{\sqrt\Delta}+{\sqrt\Delta}) A^2_{\mu} -\f{C}{\sqrt\Delta} A^1_{\mu})]+\nn \\ & &
\sum^2_{\mu,\nu=1}\bigg[e^{-\phi/2}F\f{1}{\sqrt\Delta} (A^1_{\mu}-CA^2_{\mu})(A^1_{\nu}-CA^2_{\nu})
+e^{-\phi/2}F {\sqrt\Delta} A^2_{\mu} A^2_{\nu}+
\f{e^{\phi/6}}{F^{1/3}}g_{\mu\nu}\bigg]dx^{\mu} 
dx^{\nu}\nn \\
ds^{{\prime}^2}_E&\rightarrow&\f{1}{(\ell_1\ell_6-\ell^2_3)^{3/4}}[(\f{C^2}{\sqrt\Delta}+{\sqrt\Delta})(d\varphi^2)^2+2d\varphi^2  dx^{\mu}
{\sqrt\Delta} (-\gamma B_{1\mu}+A^2_{\mu}+\sigma C_{1\mu}) -\f{C}{\sqrt\Delta}{\cal A}_{\mu}]+\nn \\& &\sum^2_{\mu,\nu=1}\bigg[\f{1}{(\ell_1\ell_6-\ell^2_3)^{3/4}}\bigg(\f{1}{\sqrt\Delta}{\cal A}_{\mu}{\cal A}_{\nu}+{\sqrt\Delta}(-\gamma B_{1\mu}+A^2_{\mu}+\sigma C_{1\mu})(-\gamma B_{1\nu}+A^2_{\nu}+\sigma C_{1\nu}) \bigg)+\nn \\& &(\ell_1\ell_6-\ell^2_3)^{1/4} g_{\mu\nu}\bigg]dx^{\mu} dx^{\nu}\nn \\
\eeqn

In order to get a feeling of the volumes let us take the  choice of a 
geometry like $AdS_5\times M^5$, where $M^5$ is 
a compact Sasaki-Einstein manifold  and without the second parameter of 
deformation i.e. with the following choice to fields
\be
\label{ads}
B_2=0;\quad C_2=0; \quad \chi=0;\quad \sigma=0,
\ee
then the computation of the volumes becomes a bit simpler. Then 
eq.(\ref{varphi2=0}) becomes
\beqn
V^{(3)}_3&=&2\pi \int dx^1 dx^2 {\sqrt {det g}} \f{e^{-\phi/12}F^{1/6}}{\Delta^{1/4}}\nn \\
V^{\prime^{(3)}}_3&=&2\pi \int dx^1 dx^2 {\sqrt {det g}} \f{e^{-\phi/12}F^{1/6}}{\Delta^{1/4}(1+\gamma^2F^2)^{1/8}}.
\eeqn
From this it follows that the integrand of deformed volume decreases. For the 
other case i.e. eq.(\ref{varphi1=0}), the volumes are
\beqn
V^{(3)}_3&=&2\pi \int dx^1 dx^2 {\sqrt {det g}} e^{-\phi/12}F^{1/6}(\f{C^2}{\sqrt\Delta}+{\sqrt\Delta})^{1/2} \nn \\
V^{\prime^{(3)}}_3&=&2\pi \int dx^1 dx^2 {\sqrt {det g}} e^{-\phi/12}F^{1/6}(\f{C^2}{\sqrt\Delta}+{\sqrt\Delta})^{1/2}\f{1}{(1+\gamma^2F^2)^{1/8}},
\eeqn
and we have the same conclusion about the integrand.
Let us go through this kind of calculation for eq.(\ref{choice_2}) i.e. for 
the second choice of the construction of cycle and the 
volumes are for the $AdS_5\times M^5$ type geometry
\beqn
V^{(2)}_3&=&\int dx^1 dx^2 dx^3 {\sqrt {det g}} e^{\phi/4} F^{-1/2}\nn \\
V^{\prime^{(2)}}_3&=& \int dx^1 dx^2 dx^3  {\sqrt {det g}} e^{\phi/4} F^{-1/2}
(1+\gamma^2F^2)^{3/8}.
\eeqn

Note that $(1+\gamma^2F^2)$ comes with a positive power. Which has got very 
interesting consequences.

In the limit eq.(\ref{ads}), it is easy to conclude whether the  masses of 
KK modes should change or not and if changes whether should increase or 
decrease and in which cases it should happen. 

The masses  of KK modes are related to the volumes of various cycles 
and for the $i$th 3-cycle it is defined as 
\be
{M^{(i)}}^3_{KK}=\f{1}{V^{(i)}_3}.
\ee 
From this definition it follows that 
\beqn
\label{def_mass}
{M^{\prime(i)}}^3_{KK}-{M^{(j)}}^3_{KK} &=& \f{V^{(j)}_3-V^{\prime(i)}_3 }{V^{(j)}_3~V^{\prime(i)}_3 }\nn \\
&=& \f{\int[f^{(j)}-f^{\prime(i)}]}{[\int f^{(j)}][\int f^{\prime(i)}]},
\eeqn
where
\be
V^{(i)}_3=\int f^{(i)};\quad V^{\prime(i)}_3=\int f^{\prime(i)}.
\ee
So, by looking at the numerator of eq.(\ref{def_mass}), we can conclude what 
happens to the masses of KK modes i.e whether the masses increases, 
decreases or
do not changes depending on eq.(\ref{def_mass}) for $>0, <0, =0$, respectively. 
In order to evaluate  eq.(\ref{def_mass}), let us assemble all the volumes for
all choices. It is easy to distinguish one volumes from the other by looking
at their superscript index, e.g. $V^{(1)}_3$ says it has come from first choice
i.e. $T~||~{\cal C}$ and similarly for others. Since, the third choice has 
got two volumes in the deformed and undeformed theory, means, we have to write
down explicitly the coordinate choice for that volume.   More importantly, 
we shall evaluate the volumes in the limit  eq.(\ref{ads})
\beqn
\label{total_vol}
V^{(1)}_3&=&(2\pi)^2 \int e^{-5\phi/12} F^{5/6}{\sqrt g_{11}} dx^1;\quad
V^{\prime^{(1)}}_3= (2\pi)^2 \int {{\sqrt g_{11}}} \f{e^{-5\phi/12} F^{5/6}} {(1+\gamma^2 F^2)^{5/8}} dx^1\nn \\
V^{(2)}_3&=&\int dx^1 dx^2 dx^3 {\sqrt {det g}} e^{\phi/4} F^{-1/2};\quad
V^{\prime^{(2)}}_3= \int dx^1 dx^2 dx^3  {\sqrt {det g}} e^{\phi/4} F^{-1/2}
(1+\gamma^2F^2)^{3/8}\nn \\
V^{(3)}_{3,\varphi^1=0}&=&2\pi \int dx^1 dx^2 {\sqrt {det g}} e^{-\phi/12}F^{1/6}(\f{C^2}{\sqrt\Delta}+{\sqrt\Delta})^{1/2} \nn \\
V^{\prime^{(3)}}_{3,\varphi^1=0}&=&2\pi \int dx^1 dx^2 {\sqrt {det g}} e^{-\phi/12}F^{1/6}(\f{C^2}{\sqrt\Delta}+{\sqrt\Delta})^{1/2}\f{1}{(1+\gamma^2F^2)^{1/8}}\nn \\
V^{(3)}_{3,\varphi^2=0}&=&2\pi \int dx^1 dx^2 {\sqrt {det g}} \f{e^{-\phi/12}F^{1/6}}{\Delta^{1/4}}\nn \\
V^{\prime^{(3)}}_{3,\varphi^2=0}&=&2\pi \int dx^1 dx^2 {\sqrt {det g}} \f{e^{-\phi/12}F^{1/6}}{\Delta^{1/4}(1+\gamma^2F^2)^{1/8}}.
\eeqn

Eq.(\ref{def_mass}) and eq.(\ref{total_vol}) for $i=j$, implies
\beqn
f^{(1)}-f^{\prime(1)}&=&{\sqrt g_{11}} e^{-5\phi/12} F^{5/6}\bigg[1-
\f{1}{(1+\gamma^2F^2)^{5/8}}\bigg] \\
f^{(2)}-f^{\prime(2)}&=&{\sqrt g} e^{\phi/4} F^{-1/2}\bigg[1-
(1+\gamma^2F^2)^{3/8}\bigg] \\
f^{(3)}_{\varphi^1=0}-f^{\prime(3)}_{\varphi^1=0}&=&{\sqrt g} e^{-\phi/12} F^{1/6} (\f{C^2}{\sqrt\Delta}+{\sqrt\Delta})^{1/2}\bigg[1-
\f{1}{(1+\gamma^2F^2)^{1/8}}\bigg] \\
f^{(3)}_{\varphi^2=0}-f^{\prime(3)}_{\varphi^2=0}&=&{\sqrt g} e^{-\phi/12} F^{1/6} \f{1}{(\Delta)^{1/4}}\bigg[1-
\f{1}{(1+\gamma^2F^2)^{1/8}}\bigg]. \\
\eeqn

It is easy to see that only  $ f^{(2)}-f^{\prime(2)}$ is negative for both 
large
and small $(\gamma F)^2$, which means that for this case masses of the KK 
modes 
decreases where as in the other cases for $i=j$ the masses increases. There 
arises a question: Is it possible to construct a 3-cycle with non-vanishing
volumes for the second choice, in practice? The answer to this question is 
known from the
studies of giant-graviton i.e. D3-branes wrapping the $S^3$ of  $AdS_5$ 
\cite{hhi}. 
More explicitly,  the $AdS_5$ geometry in the global 
coordinate is
\be
ds^2_{Ads_5}=-dt^2 cosh^2~r+dr^2+sinh^2~r ~d\Omega^2_3, \quad {\rm with}~~ R=1
\ee

which shows the presence of non-vanishing 3-cycle, and it   means, 
the choice is realisable in practice and hence the masses of KK modes can 
decrease as well. For this case the torus stay inside the $M^5$ of 
$AdS_5 \times M^5$. We can as well construct a cycle for which the torus stay 
in the $AdS_5$ part and the 3-cycle in the $M^5$.

It is very difficult to say  the  fate of the masses of KK modes for 
$i\neq j$. To say something concretely, we need to know the exact expressions 
of quantities that appear in eq.(\ref{total_vol}) and the only way to do that 
is by studying various examples.

The same kind of analysis goes through for the 2-cycles.

\section{Volume of 5-manifold}

The computation of  volume for compact 5-manifold is done by assuming that the
10-dimensional solution has a Sasaki-Einstein piece in it and the volume of 
5-manifold from eq.(\ref{e_frame}) are
\beqn
\label{vol_5}
V^{(1)}_5&=& (2\pi)^2\int dx^1 dx^2 dx^3 {\sqrt {detg}} e^{-\phi/4} F^{1/2}\nn \\
{V^{\prime}}^{(1)}_5&=&  (2\pi)^2\int dx^1 dx^2 dx^3  {\sqrt {detg}} 
(\ell_1\ell_6-\ell^2_3)^{-3/8}
\eeqn

 and the way it has been computed is as follows. Let us take a choice 
$x^4=0=\cdots0=x^8$ such that all the metric components depends only on
$x^1,x^2,x^3$ coordinates also we assume that the torus directions stay inside
the compact 5-manifold. There are also the other logical possibilities for 
which the torus can stay completely orthogonal to the 5-volume and one of the
direction of torus can stay inside the 5-volume, which is in the same spirit
for the construction of 3-cycle.  
The ratio of the integrand of  eq.(\ref{vol_5}) 
in the limit eq.(\ref{ads})  becomes
\be
\f{d{V^{\prime}}^{(1)}_5}{dV^{(1)}_5}=\f{1}{(1+\gamma^2 F^2)^{3/8}}.
\ee
In the above said limit the integrand of $V^{\prime}_5$ decreases. Defining
an object  as $ C\sim \f{1}{V_5}$, then\footnote{Note, this object is not the
central charge in the dual field theory. The way to define central charge
is by looking at the term  $\f{1}{{\cal V}_5}$ that appear in the dimensional 
reduction of
$\f{1}{2G^2_{10}}\int d^{10}x {\sqrt{g_E}} R_E=\f{{\cal V}_5}{2G^2_{5}}\int d^{5}x {\sqrt{g_E}} R_E $.} 
we  find 
\be
C^{\prime}-C\sim \f{V^{(1)}_5-{V^{\prime}}^{(1)}_5}{{V^{\prime}}^{(1)}_5V^{(1)}_5}=\f{\int[f^{(1)}_5-{f^{\prime}}^{(1)}_5]}{[\int f^{(1)}_5][\int {f^{\prime}}^{(1)}_5]}. 
\ee
Now, the sign of the term   $[f^{(1)}_5-{f^{\prime}}^{(1)}_5]$ will provide an insight 
whether the object C  changes or not. Computing it we find
\beqn
f^{(1)}_5-{f^{\prime}}^{(1)}_5&=&\f{{\sqrt {detg}}  e^{-\phi/4} F^{1/2} }
{(1+\gamma^2 F^2)^{3/8}} \bigg[ \f{3}{8} (\gamma F)^2+\cdots , \gamma F\rightarrow 0\bigg]\nn \\
&=&\f{{\sqrt {detg}}  e^{-\phi/4} F^{1/2} }
{(1+\gamma^2 F^2)^{3/8}}\bigg[(\gamma^2 F^2)^{3/8}+\cdots, \gamma F\rightarrow{\rm Large}\bigg ]
\eeqn
which shows the positive sign, means, the volume decreases.
Let us confirm our calculations 
by studying an explicit example, namely, $AdS_5\times S^5$. The metric in
Einstein frame is
\beqn
ds^2_E&=&e^{-\phi_0/2}[ds^2_{AdS_5}+ds^2_{S^5}];\quad ds^2_{S^5}=\sum^3_{i=1}d\mu^2_i+\mu^2_i d\phi^2_i ~~{\rm with} ~~\sum^3_i\mu^2_i=1,~ R=1\nn \\
{ds^{\prime}}^2_E&=&e^{-\phi_0/2}  G^{-1/4}[ds^2_{AdS_5}+\sum^3_{i=1}(d\mu^2_i+G\mu^2_i d\phi^2_i)+\gamma^2 G\mu^2_1\mu^2_2\mu^2_3 (\sum_i d\phi_i)^2].
\eeqn

 The volumes of $S^5$ and $\beta$-transformed ${S^{\prime}}^5$are
\beqn
V^{(1)}_{S^5}&=& \pi^3  e^{-5\phi_0/4}\nn \\
{V^{\prime}}^{(1)}_{S^5}&=& e^{-5\phi_0/4}  (2\pi)^3\int^{\pi/2}_0 \int^{\pi/2}_0 d\alpha d\theta c_{\theta}c_{\alpha}s_{\theta}s^3_{\alpha}\times [1+\gamma^2 s^2_{\alpha}(c^2_{\alpha}+c^2_{\theta}s^2_{\theta}s^2_{\alpha})]^{-3/8}.
\eeqn
Computation of the latter volume is very complicated and difficult too. So, we
shall evaluate it by expanding in the  $\gamma^2\rightarrow 0$ limit
and keeping terms to quadratic in $\gamma^2$ 
\be
{V^{\prime}}^{(1)}_{S^5}= e^{-5\phi_0/4}(2\pi)^3[1/8-3\gamma^2/256].
\ee
From these it follows that ${V^{\prime}}^{(1)}_{S^5}-V_{S^5}^{(1)} < 0$. 
The same is also true in the other limit i.e. when  
$\gamma^2\rightarrow {\rm Large}$. 

This conclusion of increase of the object C or decrease of the volume 
can also be seen in string 
frame. The transformed volume of $S^5$ in string frame to quadratic order in 
$\gamma$ in the $\gamma^2\rightarrow 0 $ limit is
\beqn
{V^{\prime}}^{(1)}_{S^5}&=& (2\pi)^3 [1/8-\gamma^2/32]\nn \\
V^{(1)}_{S^5}&=& \pi^3.
\eeqn  

The form of the metric and the volume  for the other choice when the torus 
stay outside the compact 5-volume is written down in eq.(\ref{choice_2}) 
with the only modification is that the sum now runs from $1 ~{\rm to}~5$ and 
$x^6=0=x^7=x^8=\varphi^1=0=\varphi^2$. These  volumes in that general 
form of geometry is not that  illuminating because we can't evaluate them. 
But in the limit, eq.(\ref{ads}), we can draw some conclusions.

For the second choice i.e. $T\perp C_5$, where $C_5$ is some kind of 
``5-cycle,''  we see that the volume increases. To see it, let us write 
$V_5=\int f_5$, then  we find
\be
f^{\prime}_5-f_5={\sqrt g} e^{5\phi/12} F^{-5/6} (1+\gamma^2F^2)^{5/8}~ >~0.
\ee  
To support this, let us take the example of $AdS_5\times S^5$ and writing
down the geometry in Einstein frame, we have 
\be
ds^2_E=e^{-\phi_0/2}[-dt^2 c^2_r +dr^2+s^2_r \sum^2_{i=1}(d\mu^2_i+\mu^2_i d\phi^2_i)+ds^2_{S^5}]~~{\rm with} \sum_i \mu^2_i=1,
\ee
where $c_r=cosh~r $ and $s_r=sinh~r$. The AdS part of the geometry is written 
in
global coordinates. Going through the $\beta$-transformation procedure, we 
find transformed geometry as
\be
{ds^{\prime}}^2_E=e^{-\phi_0/2} G^{-1/4}[-dt^2 c^2_r +dr^2+s^2_r \sum^2_{i=1}(d\mu^2_i+G\mu^2_i d\phi^2_i)+ds^2_{S^5}],
\ee
where $G^{-1}=1+4\gamma^2\mu^2_1\mu^2_2$. It is easy to see that indeed the 
transformed volume of $S^5$ increases.   
 
Similarly for the 3rd choice i.e. when the torus shares one of its direction
with the compact 5-volume then the $V_5$ decreases.  

\section{$\beta$-deformation is marginal}

It has been argued in \cite{oljm} that the $\beta$-deformation is marginal. 
In fact, we show it explicitly that indeed it is a marginal deformation by 
calculating the central charge and the dimension of operators or 
the R-charges in the gravitational side. The proof is very 
simple for the first choice, where the torus stays in the directions of 
compact 5-volume. Let us start to evaluate the following object for eq.(\ref{e_frame})
\be
\label{gr}
I\equiv \f{1}{2G^2_{10}}\int {\sqrt {g_E}} R_E,
\ee
where $g_E$ is the determinant of the 10 dimensional metric in Einstein frame
and $R_E$ is the Ricci-scalar in that frame. Evaluating  ${\sqrt{g_E}} R_E$ 
and ${\sqrt{g^{\prime}_E}} R^{\prime}_E$ and reducing it we 
find
\be
\label{central_charge1}
\f{1}{2G^2_{10}}\int d^{10}x {\sqrt {g_E}} R_{E}=\f{(2\pi)^2}{2G^2_{10}}\int d^8 x\bigg[ {\sqrt {g_E}}  R_E+\cdots \bigg] 
\ee
\be
\f{1}{2G^2_{10}}\int d^{10}x {\sqrt {g^{\prime}_E}} R^{\prime}_{E}=\f{(2\pi)^2}{2G^2_{10}}\int d^8 x\bigg[ {\sqrt {g_E}}  R_E +\cdots \bigg],         
\label{central_charge2}
\ee
where ${\sqrt{g_E}}$ and  $ R_E$ in the RHS of eq.(\ref{central_charge1}) and
 eq.(\ref{central_charge2}) are defined with respect to  the eight 
dimensional spacetime metric $g_{\mu\nu}$
transverse to the torus directions as written in eq.(\ref{e_frame}). 
The ellipses are terms coming from taking  covariant derivatives 
and  various powers of it on functions that appear in the geometry. But, we are
not interested in those terms.  The volume form for both deformed and 
undeformed metrics are
\beqn
 d^{10}x {\sqrt {g_{E,10}}}&=&  {\sqrt {g_{E,8}}} e^{\phi/6} F^{-1/3} d\varphi^1\w d\varphi^2\w dx^{1}\w\cdots\w dx^{8},\nn \\
 d^{10}x {\sqrt {g^{\prime}_{E,10}}}&= & {\sqrt {g_{E,8}}} 
(\ell_1\ell_6-\ell^2_3)^{1/4} d\varphi^1\w d\varphi^2\w dx^{1}\w\cdots\w dx^{8},
\eeqn
where as the Ricci-scalars are 
%\beqn
%ds^2&=&f_1 d\psi^2_1+f_2 d\psi^2_2+f_3 \sum^8_1 g_{\mu\nu} dx^{\mu}dx^{\nu}\\
%{ds^{\prime}}^2&=&g_1 d\Psi^2_1+g_2 d\Psi^2_2+g_3 \sum^8_1 g_{\mu\nu} dx^{\mu}dx^{\nu},
%\eeqn  
%where $f_i, g_j$ are functions of $x^{\mu}$ and define ${\tilde g}_{\mu\nu}=
%l_3 g_{\mu\nu}$ with $l_3=f_3, g_3$. Computing the Ricci scalar, we get 

\be
R_{10}=e^{-\phi/6} F^{1/3} R_8+\cdots; \quad R^{\prime}_{10}=(\ell_1\ell_6-\ell^2_3)^{-1/4} R_8+\cdots.
\ee

Dimensionally reducing  these actions to 
desired spacetime dimensions gives us interesting quantity, the central 
charge. In the 
conformal case with the geometry of the form $AdS_5\times M^5$ and 
reducing it to  5-spacetime 
dimension  i.e. we have to compactify three more directions of  
eq.(\ref{central_charge1}) and eq.(\ref{central_charge2})  
to get central charges, which are found to be same. Doing the  reduction 
once more,
from 8-dimensional spacetime to 7-dim  of eq.(\ref{central_charge1})
and eq.(\ref{central_charge2}), we get the dimension of the operators
or the R-charges. Obviously,  the reduction has to be done on the 
compact Sasaki-Einstein manifold. 

Note, that the eq.(\ref{central_charge1}) and eq.(\ref{central_charge2})  
are derived for a general case and we see that the result is  independent of
whether the undeformed theory is   conformal or non-conformal theory. 

To make things clear let us write down the   eq.(\ref{gr}) after
compactification as 
\beqn
\label{cr}
I_1&=&\f{{\cal V}_5}{2G^2_{5}} \int d^5 x \bigg[ {\sqrt {g_{5,E}}}  R_{5,E}+\cdots \bigg]  \\
I_2&=&\f{{\cal V}_3}{2G^2_{7}} \int d^7 x \bigg[ {\sqrt {g_{7,E}}}  R_{7,E}+\cdots \bigg],
\eeqn
where the quantities ${\cal V}_5, {\cal V}_3$ are related to central charge 
and the dimension of the operators or the R-charges in the dual field theory. 
The central charge is defined as 
$(c)=\f{ {\rm constant}} {{\cal V}_5}$, whereas the dimension  \cite{gk}
$\Delta=\pi\f{ N}{2} \f{Vol(\Sigma_3)}{Vol(\Sigma_5)}$. This in our notation 
becomes $\Delta= {\rm constant} \f{{\cal V}_3}{{\cal V}_5}$.

 We see that both  eq.(\ref{central_charge1}) and 
eq.(\ref{central_charge2}) have the same central charge and the dimension of
the operators as the RHS of
these two equations are same, of course without the ellipses.

The marginal behavior associated to $\beta$ defromations is shown in the 
gravity side which by AdS/CFT means its a check in the strong coupling side to
the dual field theory.  

\section{confinement} 
If we start with a background  which shows confinement at IR
then the tension of the fundamental string whose one end is fixed at infinity
possibly on a brane and the other end  probing the IR region  in the 
undeformed theory is
\be
T_s\sim  \f{e^{\phi/6}}{F^{1/3}}{\sqrt {|g_{tt}g_{xx}|}},
\ee  
where the string is stretched along the  x axis. After marginal deformation the
tension of the fundamental string goes as 
\be
T^{\prime}_s\sim (\ell_1\ell_6-\ell^2_3)^{1/4} {\sqrt {|g_{tt}g_{xx}|}}.
\ee

In general these two expressions are not same, if this is the correct way 
to find the tension of the flux tube.
To see whether the tension increases or decreases at IR. 
Let us take an example with a  choice $ \sigma=0$, keeping in mind 
the regularity of the transformed solution \cite{oljm}, and $\chi=0$  
for KS solution. This means 
\be
\label{tens}
\ell_1\ell_6-\ell^2_3=e^{2\phi/3}F^{-4/3}[(1+\gamma B_{12})^2 +\gamma^2 F^2].
\ee

At IR of KS solution \cite{ks} $B_{12}=0$, as 
\beqn
B_{12}&=&\f{g_s M}{2} s_1 s_2 s_{\psi} [f(\tau)-k(\tau)],\nn \\
f(\tau)&=& \f{(\tau coth~\tau-1)}{2sinh~\tau}[cosh~\tau-1];\quad k(\tau)=\f{(\tau coth~\tau-1)}{2sinh~\tau}[cosh~\tau+1],
\eeqn
but the second term of eq.(\ref{tens}) is non-zero and positive which means
the tension  increases. 
  
To find the energy versus distance relation we just have to integrate these
tensions over the $x$ coordinate by taking the upper limit of integration
as $x_c$, a cutoff where the end of the probe string is fixed. 

%To see whether the tension increases or decreases at IR. By  keeping terms 
%to quadratic in $\gamma^2$, we find that it increases.
  
\section{Conclusion}

There are some interesting outcomes of Lunin-Maldacena's $\beta$-deformation
technique: it can give us non-trivial potential, applying it to flat space 
yields a background reminiscent of NS5-brane background, KK modes can be
integrated out depending the choice of cycles, supersymmetry can be broken. 

The construction of cycles is done by considering the position of the torus
in the 10-dim geometry, which results in 3 possible ways to define them.
For a choice for which the torus stay outside the cycle, the transformed
volume increases and hence the masses of KK modes decreases and in all other
cases the transformed volume decreases. So, the masses of KK modes increases.  
The results of the volumes can be summarized in the following table for 
eq.(\ref{choice}). However, we expect that this behavior of volume of the 
$\beta$-transformed geometry is not going to change for other geometries.\\

\begin{tabular}{|l|l|l|l|c|}
\hline
\multicolumn{1}{|c|}{ All cycles} &
\multicolumn{1}{c|}{ Torus $||$ $ Cycles (C_i)$} &
\multicolumn{1}{c|}{ Torus $\perp ~Cycles (C_i)$} &
\multicolumn{1}{c|}{ Torus $\cap ~ Cycles (C_i)$} &\\
\hline
$V^{\prime}_i-V_i$ & Decreases & Increases & Decreases &\\
& (negative) & (positive) & (negative) &\\
\hline
$M^{\prime}_{KK}-M_{KK}$ & Increases & Decreases & Increases &\\
& (positive) &(negative) &(positive) &\\
\hline
\end{tabular}\\

It would be interesting to see whether there exists any confining geometry
 for which the volume increases, if they are there then it does 
not looks like we can get a pure super Yang-Mills at IR.

In general, we know that the presence of non-zero fluxes to NSNS and 
RR 3-form field strength break conformal invariance. But,  this way of 
generating 
$\beta$-transformation do not in fact care whether these 3-form field 
strengths 
are zero or non-zero in the sense that the term ${\cal V}_5$ and ${\cal V}_3$ 
that appear in eq.(\ref{cr})  do not depends whether  we are 
dealing with a conformal or a non-conformal theory.

\section{Acknowledgment}
I would like to thank O. Aharony, D. Gepner and J. Sonnenschein
 for useful discussions and
 Oleg Lunin for many useful correspondences. The financial help
from Feinberg
graduate school  is gratefully acknowledged.

\end{document}